\documentclass[12pt]{iopart}

\usepackage{iopams}
\usepackage[amssymb]{SIunits}
\usepackage{graphicx}
\begin{document}

\title[Towards electron transport measurements in
chemically modified graphene]{Towards electron transport
measurements in chemically modified graphene: The effect of a
solvent}

%\author{Arnhild Jacobsen \footnote{Both authors have contributed equally to this
%work.}, Fabian M. Koehler \footnotemark[1], Wendelin J. Stark and
%Klaus Ensslin}

\author{Arnhild Jacobsen \footnote{Both authors have contributed equally to this work.}}
\address{Solid State Physics
Laboratory, ETH Zurich, Switzerland}
\ead{arnhildj@phys.ethz.ch}

\author{Fabian M. Koehler \footnotemark[1]}
\address{Institute for Chemical and Bioengineering, ETH Zurich, Switzerland}
\ead{fabian.koehler@chem.ethz.ch}

\author{Wendelin J. Stark}
\address{Institute for Chemical and Bioengineering, ETH Zurich, Switzerland}

\author{Klaus Ensslin}
\address{Solid State Physics
Laboratory, ETH Zurich, Switzerland}

%%%%%%%%%%%%%%%%%%%%%%%%%%%%%%%%%%%%%%%%%%

\begin{abstract}
Chemical functionalization of graphene modifies the local electron
density of the carbon atoms and hence electron transport. Measuring
these changes allows for a closer understanding of the chemical
interaction and the influence of functionalization on the graphene
lattice. However, not only chemistry, in this case diazonium
chemistry, has an effect on the electron transport. Latter is also
influenced by defects and dopants resulting from different
processing steps. Here, we show that solvents used in the chemical
reaction process change the transport properties. In more detail,
the investigated combination of isopropanol and heating treatment
reduces the doping concentration and significantly increases the
mobility of graphene. Furthermore, the isopropanol treatment alone
increases the concentration of dopants and introduces an asymmetry
between electron and hole transport which might be difficult to
distinguish from the effect of functionalization. The results shown
in this work demand a closer look on the influence of solvents used
for chemical modification in order to understand their influence.

\end{abstract}

%%%%%%%%%%%%%%%%%%%%%%%%%%%%%%%%%%%%%%%%%%%%%%%

\maketitle

%\tableofcontents

%%%%%%%%%%%%%%%%%%%%%%%%%%%%%%%%%%%%%%%%%%%%%
%Introduction
%%%%%%%%%%%%%%%%%%%%%%%%%%%%%%%%%%%%%%%%%%%%%

\section{Introduction}

Graphene is an electronic material with high electron mobilities
even at room temperature\cite{geim2007}. Usually graphene is
prepared by exfoliating individual layers from bulk graphite and
putting them down on a substrate\cite{novoselov2004}. With such
techniques it has become possible to prepare samples displaying
quantum Hall effect, testifying to the high electronic quality of
such systems\cite{novoselov2005,zhang2005}. Further improved
mobilities were achieved by suspending graphene
flakes\cite{bolotin2008}, or, very recently, by depositing graphene
on boron nitride\cite{dean2010}. It is generally believed that
unintentional adatoms on top of the graphene flake and charge traps
in the substrate limit the mobility for conventional
devices\cite{chen2008,chen2008b,schedin2007}.

Chemical modification of graphene has been achieved by a number of
methods and has been investigated by Raman measurements and
transport studies\cite{farmer2009,sinitskii2010,koehler2010,
sharma2010}. Applying chemistry on graphene changes the local
carbon-carbon bond structure, the orbitals and hence the electronic
properties of the material. Until now, it is not so clear how
conventional methods used in almost any graphene sample preparation,
such as baking in inert gas atmosphere in combination with rinsing
in water or organic solvents affect the electronic quality of a
graphene system. Such treatments are also standard conditions in
chemical reactions and can induce a change on electron transport
along with chemical functionalization itself. Therefore solvent
effects should be taken into account when analyzing transport data
of chemically derivatized graphene samples.

In the first part of this work we present a confocal Raman
spectroscopy analysis of graphene chemically modified with aromatic
diazonium ions. A difference in reactivity between single layer,
bi-layer and single layer edge is observed. In the second part we
first show the influence of functionalization on the electronic
transport properties of graphene, and afterwards we focus on the
influence of repeated treatment with baking and rinsing in
isopropanol. Here we find that the treatment with only isopropanol
leads to an increase in the doping concentration and an asymmetry
between electron and hole transport which is partly similar to the
effect of the functionalization. In addition we observe that the
combined treatment with isopropanol and baking leads to a higher
electronic quality than just heating alone. This is further
investigated at low temperatures in the last part of this paper.

%%%%%%%%%%%%%%%%%%%%%%%%%%%%%%%%%%%%%%%%%%%%%%%%%%%%%%%%
% Experimental methods
%%%%%%%%%%%%%%%%%%%%%%%%%%%%%%%%%%%%%%%%%%%%%%%%%%%%%%%%

\section{Experimental method}
\label{exp method}

Single and bi-layer graphene flakes were exfoliated from natural
graphite and deposited onto a Silicon substrate covered by
$\approx$285\,nm thermal silicon dioxide\cite{novoselov2004} and
identified using Raman spectroscopy and light
microscopy\cite{ferrari2006,graf2007}.

For the Raman spectroscopy study the chemical functionalization is
carried out at room temperature by immersing the chip into a
20\,mmolL$^{-1}$ solution of water-soluble nitrobenzene diazonium
salt (4-nitrobenzene diazonium tetrafluorborate from Sigma
Aldrich)\cite{koehler2010}. After the functionalization the chips
were cleaned once in isopropanol (1 min), two times in water (1
min), a second time with isopropanol (1 min) and finally blown dry
with nitrogen. For the electronic transport experiments the chemical
functionalization was carried out at $0\degree$C using a
4\,mmolL$^{-1}$ solution. The cleaning procedure after the
functionalization was the same as for the Raman spectroscopy study.
It should be pointed out that due to the different reaction
conditions described above the Raman spectroscopy study and the
transport study cannot be directly compared in terms of amount of
induced disorder as a function of reaction time.

During the Raman spectroscopy study laser power was kept at 2\,mW in
order to avoid heating and the introduction of defects due to the
laser.

For the electronic transport experiments Ohmic contacts were defined
on the graphene flakes using standard electron beam lithography
techniques followed by the evaporation of Cr/Au (2/40\,nm). The
highly doped silicon substrate is used as a global gate to tune the
overall Fermi energy of the device. The Hall bar used to investigate
the influence of isopropanol and heating treatment on the transport
properties of graphene is patterned in a second electron beam
lithography step followed by reactive ion etching.

%%%%%%%%%%%%%%%%%%%%%%%%%%%%%%%%%%%%%%%%%%%%%%%%%%%%%%%%%%%%%%%%%%%
%%%%%%%%%%%%%%%%%%%%%%%%%%%%%%%%%%%%%%%%%%%%%%%%%%%%%%%%%%%%%%%%%%%%%

\section{Results and discussion}

%%%%%%%%%%%%%%%%%%%%%%%%%%%%%%%%%%%%%%%%%%%%%%%%%%%%%%%%
% Characterisation at 0T
%%%%%%%%%%%%%%%%%%%%%%%%%%%%%%%%%%%%%%%%%%%%%%%%%%%%%%%%

\subsection{Raman spectroscopy of chemically functionalized graphene}

\begin{figure}
  \begin{center}
    \includegraphics[width=0.8\textwidth]{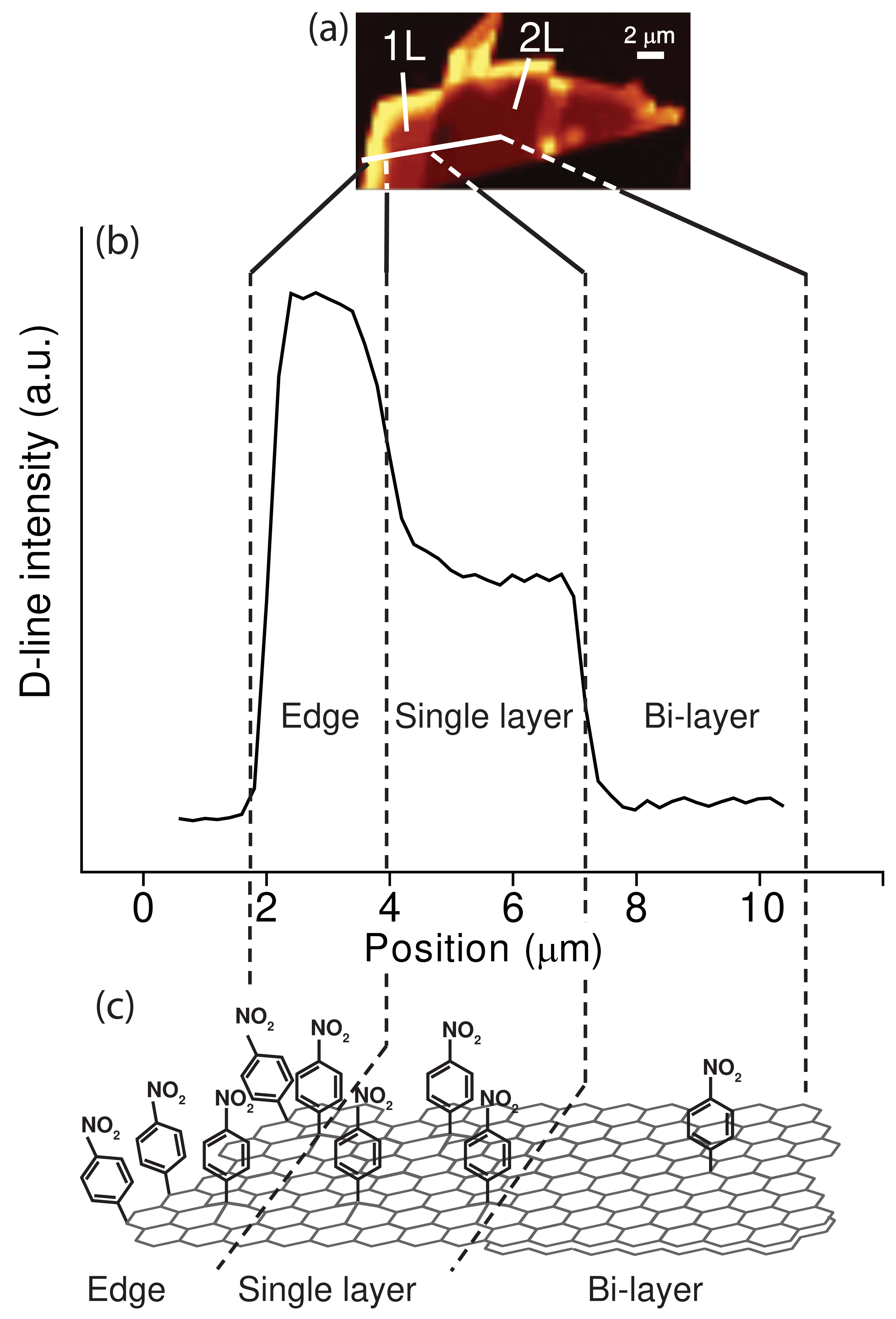}
    \caption{(a) Raman 2D map of the D-line intensity of the investigated graphene flake.
    (b) D-line intensity along the white line in (a). (c) Schematic representation
    of the difference in reactivity between the edge area, the single layer area and the bi-layer area.}
    \label{fig1}
  \end{center}
\end{figure}

Raman spectroscopy is a powerful tool for both identifying the
number of graphene layers\cite{ferrari2006,graf2007} as well as for
monitoring doping\cite{casiraghi2007, stampfer2007},
defects\cite{dresselhaus2010} and chemical
functionalization\cite{koehler2010,sharma2010} of graphene. The most
prominent features in the Raman spectra of graphene are the G band
(around 1580 cm$^{-1}$) and the 2D band (around 2700 cm$^{-1}$). In
addition, in the presence of defects or at the edge of graphene the
disorder induced D-line located around 1350\,cm$^{-1}$ can be
observed\cite{malard2009}. Here we functionalize graphene using
diazonium chemistry and monitor the introduction of defects in
graphene lattice by measuring the intensity of the D-line of the
Raman spectra.

Diazonium chemistry has previously been used to functionalize a
variety of carbon forms
\cite{usrey2005,koehler2009,grass2007,rossier2009,koehler2009b} and
it was recently shown that also graphene can be functionalized in a
similar manner as the other carbons forms using the same
chemistry\cite{koehler2010,sharma2010}. In this experiments a flake
is used that has both single and bi-layer domains, allowing the
direct identification of differences in the chemical reactivity
towards the diazonium reagent. In Fig.\,\ref{fig1}(a) the 2D map of
the integrated D-peak intensity is shown after 20 minutes immersion
in the reaction medium. Bright areas correspond to high intensity
and dark areas to low intensity. Three distinct domains with similar
intensities are visible, which can be attributed to bi-layer, single
layer or single layer edge of the graphene flake, respectively. This
is further shown in Fig.\,\ref{fig1}(b) where the intensity of the
D-peak along the white line in Fig.\ref{fig1}(a) is plotted.

The D-peak intensity of the Bi-layer region is very hard to
identify, as the signal overlaps with adsorbed
species\cite{koehler2010}. It has been shown by Strano et
al.\cite{sharma2010} that long reaction time and extensive washing
procedure is necessary to identify a small D-peak on Bi-layer
graphene after functionalization with diazonium ions. On single-
layer graphene the D-line integral is significantly higher than on
bi-layer. The higher reactivity towards diazonium chemistry for
single-layer than bi-layer has been attributed to less ripples on
the bi-layer surface\cite{koehler2010} and to screening of
electron-hole puddles in bi-layer\cite{sharma2010}. Furthermore the
single layer part can be divided in two regions of distinct
intensities, bulk single layer with a lower D-peak intensity and
edge single layer with a higher intensity. This increased edge
intensity was also shown earlier and is attributed to a higher
reactivity of the reagents towards the edge, due to a higher degree
of flexibility, which is necessary to change the local geometry from
planar sp$^2$ to tetrahedral sp$^3$. In addition it was recently
shown that these edge regions grow over the whole single layer area
with prolonged reaction time\cite{koehler2010}. This is an
indication that near defects or functional groups on the surface,
the carbon atoms react more easily with the diazonium reagents. In
Fig.\ref{fig1}(c) the difference in reactivity between the different
parts of the flake is schematically illustrated. For a detailed
investigation of the dependence of disorder as function of exposure
time to the reaction medium see Koehler et.al\cite{koehler2010}.

%%%%%%%%%%%%%%%%%%%%%%%%%%%%%%%%%%%%%%%%%%%%%%%%%%%%%%%%
% Room temperature transport measurements
%%%%%%%%%%%%%%%%%%%%%%%%%%%%%%%%%%%%%%%%%%%%%%%%%%%%%%%%

\subsection{Room temperature transport measurements}

The possibility of controlled doping and, as investigated above,
selective functionalization of graphene edges makes chemical
modification of graphene interesting for electronic transport
experiments.

\begin{figure}
  \begin{center}
    \includegraphics[width=0.8\textwidth]{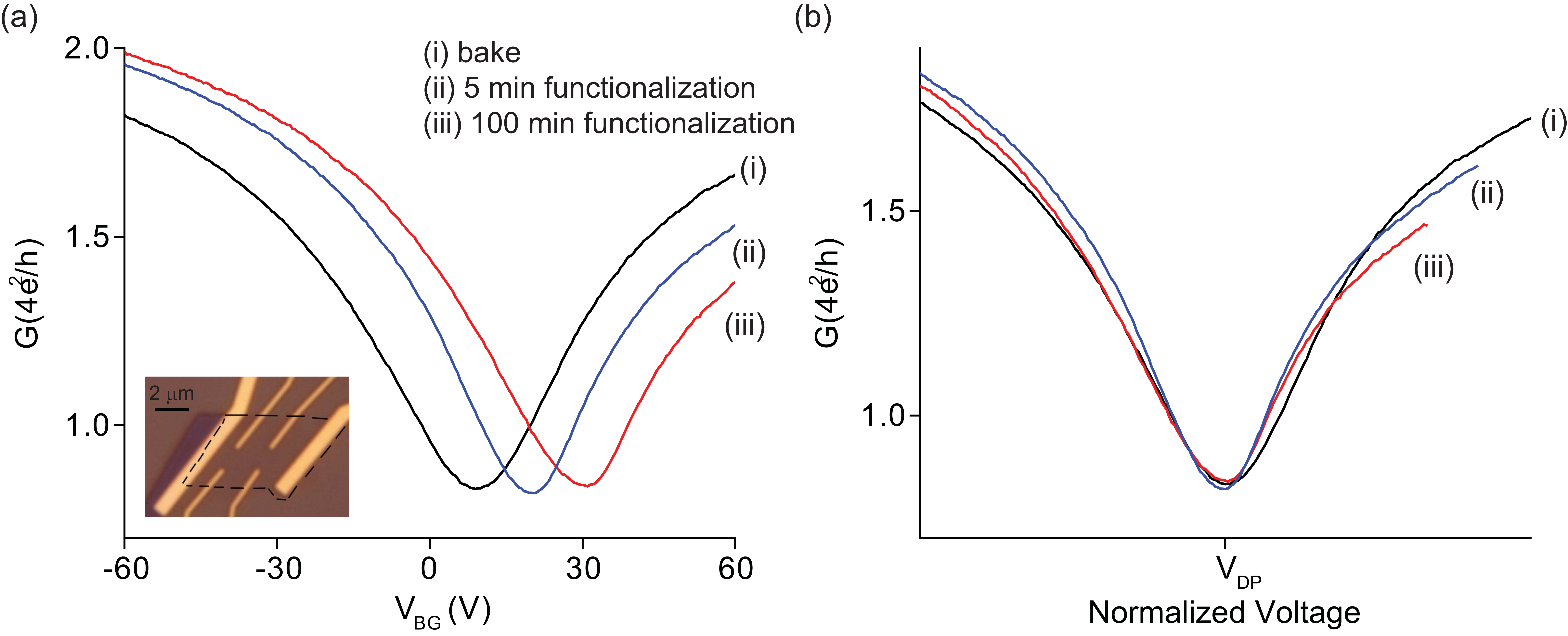}
    \caption{(a) Room temperature measurement of the two-terminal conductance as a function of backgate voltage (i) after baking the sample,
    (ii) after 5 minutes functionalization and (iii) after 100 minutes functionalization.
    (b) The curves in (a) laterally shifted and normalized with respect to the Dirac point ($V_\mathrm{DP}$)
    The inset in (a) shows a light microscope image of the investigated sample where the dotted line indicates the contour of the graphene flake.}
    \label{fig2}
  \end{center}
\end{figure}

In Fig.\,\ref{fig2}(a) room temperature measurements of the
conductance (G) as a function of backgate voltage ($V_\mathrm{BG}$)
are shown for (i) an unfunctionalized sample (only heated in order
to remove dopants from the surface), (ii) after 5 minutes of
functionalization and (iii) after 100 minutes of functionalization.
These are two-terminal measurements on an unpatterned graphene flake
(see the light microscope image of the measured device in the inset
in Fig.\,\ref{fig2}). From Fig.\,\ref{fig2}(a) it can clearly be
seen how the functionalization leads to an increased p-doping of the
graphene flake. Before functionalization the point of minimum
conductance (the Dirac point, $V_\mathrm{DP}$) is located at +9\,V
in backgate. After 5 minutes of functionalization the Dirac point is
shifted to +21\,V and after 100 minutes of functionalization the
Dirac point is at +31\,V. In Fig.\,\ref{fig2}(b), where the backgate
traces are normalized with respect to $V_\mathrm{DP}$, it can be
seen that the functionalization introduces a small asymmetry between
electron and hole transport. This asymmetry is much weaker than
observed previously by Farmer et al\cite{farmer2009}. In addition it
can be seen from Fig.\,\ref{fig2}(b) that the mobility (slope of G
versus voltage) of the graphene flake is not significantly changed
after functionalization. Both observations can be explained by a
lower amount of functionalization due the low temperature
($0\degree$C) used in this work.

In the functionalization process described above the graphene flake
is first immersed in water containing the reactive diazonium ions
and afterwards in isopropanol to remove unreacted species and
improve the drying step. In order to analyze the influence only of
the chemical functionalization on the electronic transport
properties of graphene it is crucial to know the effect of the
involved solvents. In the following we will therefore investigate
the influence of isopropanol and baking on graphene's transport
properties, which are part of  the chemical and physical treatments
involved in the reaction and measurement process.

\begin{figure}
  \begin{center}
    \includegraphics[width=0.8\textwidth]{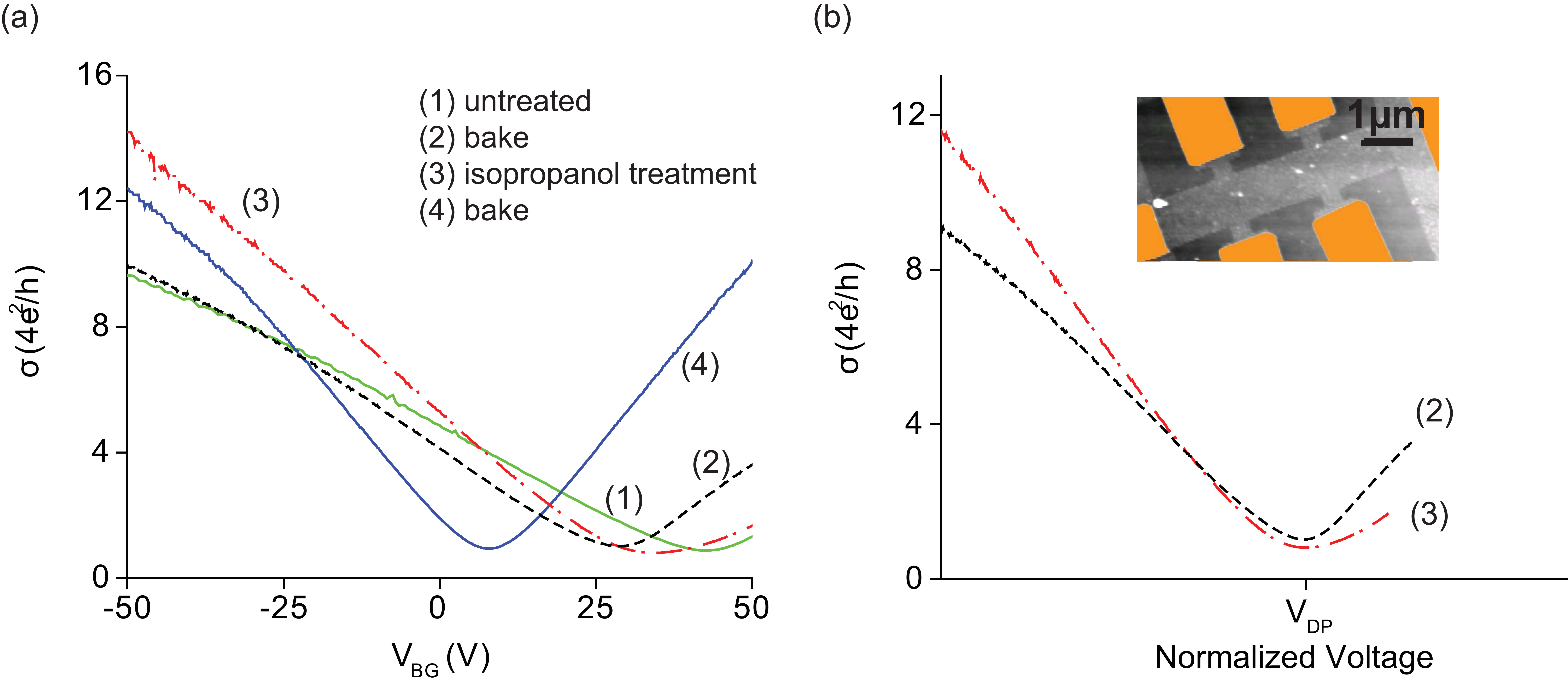}
    \caption{(a) Room temperature measurement of the four-terminal conductivity as a function of backgate voltage for (1) an untreated Hall
    bar, (2) after heating, (3) after isopropanol treatment and (4) after a second heating. (b) The curves after (2) baking and (3) isopropanol treatment from (a)
    normalized with respect to the Dirac point ($V_\mathrm{DP}$).
    The inset shows a scanning force micrograph (SFM) image of the measured Hall bar where the contacts hare highlighted in orange for clarity.}
    \label{fig3}
  \end{center}
\end{figure}

To investigate the influence of isopropanol and heating we use the
Hall bar shown in the inset in Fig.\,\ref{fig3}(a). The width of the
Hall bar is $\approx$1\,$\mu$m and the length between two voltage
probes is $\approx$2\,$\mu$m. All following measurements are
four-terminal measurements. For the isopropanol treatment the chip
with the Hall bar is immersed in isopropanol for 5\,minutes and
afterwards blown dry with nitrogen gas. The heating of the sample is
done in the sample holder while the vacuum is constantly pumped. In
order to monitor changes in the conductivity of the sample during
the heating a constant current of 10\,nA is applied to the Hall bar
and the four-terminal resistance is measured at
$V_\mathrm{BG}=0$\,V. The sample is always heated at 150$\degree$C
until the measured resistance is stable. This may take many hours.

In Fig.\,\ref{fig3}(a) the conductivity ($\sigma$) of the Hall bar
as a function of applied V$_{\mathrm{BG}}$ for (1) the untreated
sample, (2) after heating the sample, (3) after treating the sample
with isopropanol and (4) after heating the sample again is plotted.
It can be seen that both the mobility and the position of the Dirac
point is changed significantly after the different treatments.

For the untreated sample the Dirac point is located at +43\,V. The
extensive doping of the pristine sample is probably due to resist
residues and other dopants accumulated during the processing steps.
In order to remove these dopants we always bake our samples before
starting measurements (as we also did before functionalization).
Here it can be seen that after the initial baking of the sample the
Dirac point has moved to +26\,V. The corresponding change in
mobility will be discussed below. As a next step we treat the sample
with isopropanol. Fig.\,\ref{fig3} shows that the Dirac point is
shifted from +29\,V to +34\,V in backgate after the isopropanol
treatment, which means that isopropanol significantly p-dopes
graphene. From Fig.\ref{fig3}(b), where the traces before and after
isopropanol treatment from Fig.\,\ref{fig3}(a) are normalized with
respect to $V_\mathrm{DP}$, it can in addition be seen that the
isopropanol introduces a strong asymmetry between the electron and
hole conductivities. Above it has been shown that in the absence of
significant sp$^3$ hybridization of the graphene surface,
functionalization with diazonium salt does not lead to a suppression
of conductance, only a shift of the Dirac point to more positive
backgate voltages. The observed asymmetry after isopropanol
treatment is larger than observed after the functionalization
(Fig.\,\ref{fig2}(b)). However, it is similar to the asymmetry found
by Farmer et. al after functionalization\cite{farmer2009}. The
qualitative similarities between the changes in the conductivity of
graphene after isopropanol treatment and the changes observed after
functionalization suggest that with the functionalization procedure
described above it might be difficult to separate the effects of the
diazonium salt and the effects of isopropanol.

In the final step we heat the sample a second time in order to see
if we can remove the dopants introduced by the isopropanol
treatment. Surprisingly the Dirac point does not only shift back to
+29\,V where it was located before the isopropanol treatment, it
shifts much further to +8\,V. Together with the corresponding
increase in mobility this suggests that the electronic quality of
graphene can be improved by repeated isopropanol treatments followed
by heating. In case of the measurements (1) and (3) it is difficult
to extract the electron mobilities and thus only the hole mobilities
will be compared in the following. For the untreated graphene flake
(1) we obtain a hole mobility of 2100\,cm$^{2}$/Vs. After the first
heating step (2) the mobility has increased to 2700\,cm$^{2}$/Vs.
The following isopropanol treatment (3) increases the mobility
further to 3600\,cm$^{2}$/Vs and after the last heating (4) the
mobility reaches 4700\,cm$^{2}$/Vs. Generally we expect the
introduction/removal of dopants to decrease/increase the mobility.
Here, after the isopropanol treatment, an increase in hole mobility
is observed together with an increased doping. This might be due to
the removal of some dopants and the introduction of a different kind
of dopants.

The fact that annealing the sample removes dopants and improves the
mobility is generally accepted. Therefore baking is normally a part
of standard processing procedures for graphene. However, that a
subsequent treatment with isopropanol followed by annealing is
removing even more dopants has to our knowledge not been noted so
far. We observe here that the repeated treatment with isopropanol
followed by heating improves the quality of the sample far beyond
the improvement due to the first heating.

In addition to the observed increase in sample quality after the
combination of isopropanol treatment and heating the effect of the
isopropanol treatment alone should also be pointed out. Isopropanol
treatment alone leads to an increased p-doping and electron-hole
asymmetry. These two effects are partly seen after chemical
functionalization as well and thus it is important to be very
careful when assigning shifts of the Dirac point and changes in
electron-hole symmetry solely to the introduction of the modifying
species.

In order to evaluate the connection between the functional groups
and the transport experiments an estimate of the mean distance
between the defects induced by the functional groups is necessary.
This may be possible by evaluating the Raman data as shown by
Lucchese et.al\cite{lucchese2010}. However our Raman data and
transport data are from two different measurement cycles on
different samples. Simulations showing the connection of defect
spacing and transport in graphene nanoribbons have been shown by
Lopez-Bezanilla et.al\cite{bezanilla2009}. For further investigation
of functionalized graphene and the influence of different solvents
it would be therefore be favourable to perform Raman spectroscopy
studies parallel with transport studies in order to make a more
quantitative study about the defect density.

%%%%%%%%%%%%%%%%%%%%%%%%%%%%%%%%%%%%%%%%%%%%%%%%%%%%%%%%
% Low temperature transport measurements
%%%%%%%%%%%%%%%%%%%%%%%%%%%%%%%%%%%%%%%%%%%%%%%%%%%%%%%%

\subsection{Low temperature transport measurements}

\begin{figure}
  \begin{center}
    \includegraphics[width=0.8\textwidth]{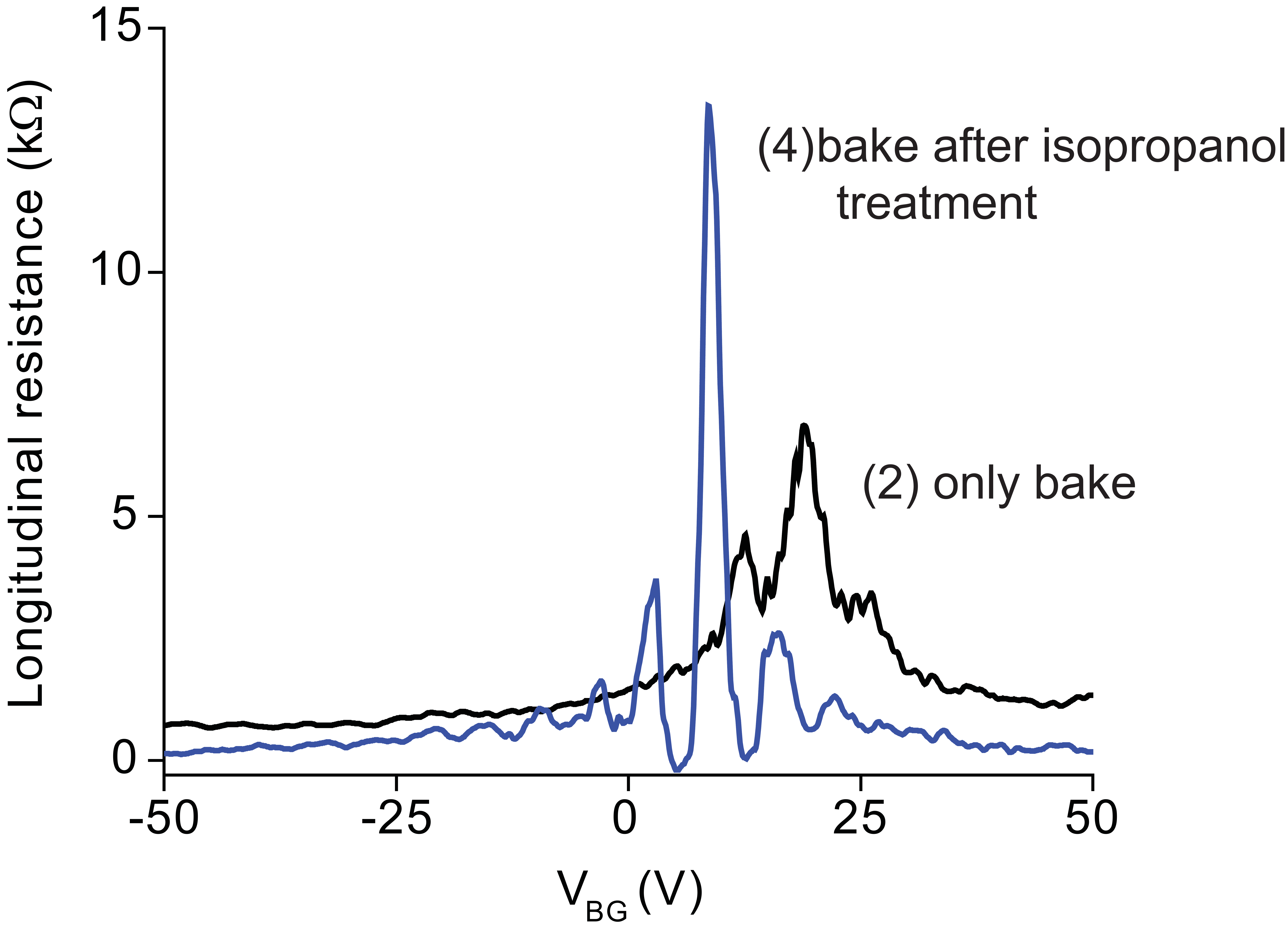}
    \caption{Longitudional resistance as a function of backgate voltage at
    4K after (2) heating and (4) isopropanol treatment and heating measured at a magnetic field $B=5$\,T.}
    \label{fig4}
  \end{center}
\end{figure}

The quantum Hall effect and the corresponding magnetooscillations of
the longitudional resistance is found in two-dimensional systems of
high quality and at low temperatures. The quality of the quantum
Hall effect is a direct measure for the quality of the electronic
system. Therefore, in order to further investigate the influence of
the isopropanol treatment and confirm the improvement of the
electronic quality of the graphene, we perform transport
measurements in magnetic field at $T=$4\,K.

Fig.\,\ref{fig4} shows the four-point longitudional resistance
($R_\mathrm{xx}$) of the flake as a function of $V_\mathrm{BG}$ at
fixed magnetic field $B=$5\,T after the first time heated (2) and
after isopropanol treatment and the second heating (4). ((2) and (4)
corresponds to Fig.\,\ref{fig3}). Before the isopropanol treatment
$R_\mathrm{xx}$ does not go to zero and only a weak splitting of the
main resistance peak is observed. In contrast, after the isopropanol
treatment and heating $R_\mathrm{xx}$ is clearly zero for filling
factor $\nu=$2 and in addition several more oscillations in
$R_\mathrm{xx}$ are visible.

These measurements show that the electronic quality of the graphene
flake is indeed improved after treating it with isopropanol and
heating it.

%%%%%%%%%%%%%%%%%%%%%%%%%%%%%%%%%%%%%%%%%%%%%%%%%%%
% Conclusion
%%%%%%%%%%%%%%%%%%%%%%%%%%%%%%%%%%%%%%%%%%%%%%%%%%%

\section{Conclusions}

To conclude we have presented confocal Raman spectroscopy studies of
chemically functionalized single and bi-layer graphene and shown
that the reactivity of the edges and the single layer parts are
larger than the reactivity of the bi-layer parts. Furthermore we
have performed a transport study of chemically modified graphene and
found that the influence of an isopropanol treatment is comparable
to the influence of the functionalization itself. It is shown that
on one hand isopropanol leads to a p-doping similar to the p-doping
observed after functionalization. In addition it is observed that
isopropanol treatment followed by heating significantly improves the
electronic quality of graphene beyond the improvement due to heating
alone.

%%%%%%%%%%%%%%%%%%%%%%%%%%%%%%%%%%%%%%%%%%%%%%%%%%%
%%%%%%%%%%%%%%%%%%%%%%%%%%%%%%%%%%%%%%%%%%%%%%%%%%%%%

\section{Acknowledgements}

The authors thank Prof. Christofer Hierold for access to the
confocal Raman microscope, T. Ihn for helpful discussions and the
Swiss National Science Foundation for financial support.

%%%%%%%%%%%%%%%%%%%%%%%%%%%%%%%%%%%%%%%%%%%%%%%%%%%
%%%%%%%%%%%%%%%%%%%%%%%%%%%%%%%%%%%%%%%%%%%%%%%%%%%%%

%%%%%%%%%%%%%%%%%%%%%%%%%%%%%%%%%%%%%%%%%%%%%%%%
%%%%%%%%%%%%%%%%%%%%%%%%%%%%%%%%%%%%%%%%%%%%%%%%%

\end{document}